\documentclass{jaa}

\usepackage{graphicx}
\usepackage{xspace}	

\usepackage{threeparttable}
\usepackage{amsmath}
\usepackage[utf8]{inputenc}
\usepackage{longtable}
\usepackage{graphicx}
\usepackage{natbib}
\usepackage{multirow}
\usepackage[Table,xcdraw]{xcolor}
\usepackage{rotating}
\newcommand{\apj}{ApJ}

\newcommand{\mnras}{MNRAS}

\newcommand{\physrep}{Physics Reports}
\newcommand{\nat}{Nature}
\newcommand{\apjl}{ApJL}
\newcommand{\apjs}{ApJS}

\newcommand{\aap}{Astronomy and Astrophysics}

\newcommand{\aj}{AJ}
\newcommand{\pasa}{PASA}
\newcommand{\ssr}{Space Science Reviews}

\newcommand{\AstroSat}{{\em AstroSat}\xspace}
\newcommand{\fermi}{{\em Fermi}\xspace}
\newcommand{\kw}{{\em Konus}-Wind\xspace}

\newcommand{\swiftT}{{T$_{\rm 0}$}\xspace}

\newcommand{\keV}{{\rm keV}\xspace}

\newcommand{\swift}{{\em Swift}\xspace}
\newcommand{\tninty}{{$T_{\rm 90}$}\xspace}

\newcommand{\Ep}{$E_{\rm p}$\xspace}
\newcommand{\sw}[1]{\texttt{#1}}
\newcommand{\af}{\sw{afterglowpy}}

\usepackage{hyperref}
\hypersetup{colorlinks=true,
    breaklinks=true,
    urlcolor= black,
    linkcolor= blue,
    bookmarksopen=false,
    filecolor=black,
    citecolor=blue,
    linkbordercolor=blue
}

\begin{document}\sloppy

\title{Revealing nature of GRB 210205A, ZTF21aaeyldq (AT2021any), and follow-up observations with the 4K$\times$4K CCD Imager+3.6m DOT}


\author{Rahul Gupta\textsuperscript{1,2,*}, Amit Kumar\textsuperscript{1,3}, Shashi Bhushan Pandey\textsuperscript{1}, A. J. Castro-Tirado\textsuperscript{4,5}, Ankur Ghosh\textsuperscript{1,3}, Dimple\textsuperscript{1,2}, Y.-D. Hu\textsuperscript{4}, E. Fern\'andez-Garc\'ia\textsuperscript{4}, M. D. Caballero-Garc\'ia\textsuperscript{4}, M. \'A. Castro-Tirado\textsuperscript{4}, R. P. Hedrosa\textsuperscript{6}, I. Hermelo\textsuperscript{6}, I. Vico\textsuperscript{6}, Kuntal Misra\textsuperscript{1}, Brajesh Kumar\textsuperscript{1}, Amar Aryan\textsuperscript{1,2}, and Sugriva Nath Tiwari\textsuperscript{2}}
\affilOne{\textsuperscript{1}Aryabhatta Research Institute of Observational Sciences (ARIES), Manora Peak, Nainital-263002, India.\\}
\affilTwo{\textsuperscript{2}Department of Physics, Deen Dayal Upadhyaya Gorakhpur University, Gorakhpur-273009, India.\\}
\affilThree{\textsuperscript{3}School of Studies in Physics and Astrophysics, Pt. Ravishankar Shukla University, Chattisgarh 492010, India.\\}
\affilFour{\textsuperscript{4}Instituto de Astrof\'isica de Andaluc\'ia (IAA-CSIC), Glorieta de la Astronom\'ia s/n, E-18008, Granada, Spain.\\}
\affilFive{\textsuperscript{5}Departamento de Ingenier\'ia de Sistemas y Autom\'atica, Escuela de Ingenier\'ias, Universidad de M\'alaga, C\/. Dr. Ortiz Ramos s\/n, 29071 M\'alaga, Spain.\\}
\affilSix{\textsuperscript{6}Centro Astron\'mico Hispano Alem\'an. Observatorio de Calar Alto. Sierra de los Filabres, E-04550, G\'ergal, Almerí\'ia, Spain.\\}

\twocolumn[{

\maketitle

\corres{rahulbhu.c157@gmail.com, rahul@aries.res.in, shashi@aries.res.in}

\msinfo{2021}{---}

\begin{abstract}
Optical follow-up observations of optical afterglows of gamma-ray bursts are crucial to probe the geometry of outflows, emission mechanisms, energetics, and burst environments. We performed the follow-up observations of GRB 210205A and ZTF21aaeyldq (AT2021any) using the 3.6m Devasthal Optical Telescope (DOT) around one day after the burst to deeper limits due to the longitudinal advantage of the place. This paper presents our analysis of the two objects using data from other collaborative facilities, i.e., 2.2m Calar Alto Astronomical Observatory (CAHA) and other archival data. Our analysis suggests that GRB 210205A is a potential dark burst once compared with the X-ray afterglow data. Also, comparing results with other known and well-studied dark GRBs samples indicate that the reason for the optical darkness of GRB 210205A could either be intrinsic faintness or a high redshift event. Based on our analysis, we also found that ZTF21aaeyldq is the third known orphan afterglow with a measured redshift except for ZTF20aajnksq (AT2020blt) and ZTF19abvizsw (AT2019pim). The multiwavelength afterglow modelling of ZTF21aaeyldq using the $\af$ package demands a forward shock model for an ISM-like ambient medium with a rather wider jet opening angle. We determine circumburst density of $n_{0}$ = 0.87 cm$^{-3}$, kinetic energy $E_{k}$ = 3.80 $\times 10^{52}$ erg and the afterglow modelling also indicates that ZTF21aaeyldq is observed on-axis ($\theta_{obs} < \theta_{core}$) and a gamma-ray counterpart was missed by GRBs satellites. Our results emphasize that the 3.6m DOT has a unique capability for deep follow-up observations of similar and other new transients for deeper observations as a part of time-domain astronomy in the future.
\end{abstract}

\keywords{gamma-ray burst: general, gamma-ray burst: individual: GRB 210205A and ZTF21aaeyldq, methods: data analysis, telescope}

}]


\doinum{}
\artcitid{\#\#\#\#}
\volnum{000}
\year{0000}
\pgrange{1--}
\setcounter{page}{1}
\lp{15}

\section{Introduction}
\label{Intro}

Gamma-ray bursts (GRBs) are among the most fascinating and luminous explosive transients, occurring at cosmological distances in the Universe. They have two distinct phases of emission; one is the prompt emission (the initial burst phase, peak at sub-MeV energy range), followed by a long-lived multiwavelength afterglow phase \citep{2015PhR...561....1K}. The prompt emission can typically last from ms to minutes, whereas afterglow lasts from days to months, sometimes even in years (radio afterglows, \cite{2012ApJ...746..156C}). Origin wise, GRBs have been classified as core-collapse or compact objects merger origins such as two neutron stars (NS) merging each other, or NS- black hole (BH), or NS-white dwarf (WD) merger \citep{2016SSRv..202...33L}. Observationally, GRBs are classified into two different classes based on the duration of prompt gamma-ray emission. One set of bursts for which duration is less than 2 s have a harder spectrum, known as short-hard burst, and the other one is long-soft burst, having durations greater than 2 s \citep{1993ApJ...413L.101K}. However, the recent discovery of a short GRB 200826A from a collapsar challenges the traditional classification of GRBs \citep{2021NatAs.tmp..142A}. 

In both the cases (long and short GRBs), we see the $\gamma$-ray emission due to jetted emission with high Lorentz factors powered by a possible central engine. The ejected material out-flowing with high Lorentz factors can have different structures, and various shells are moving with different Lorentz factors can collide with each other creating internal shock, which is responsible for the prompt emission \citep{2013FrPhy...8..661G, 2018JApA...39...75I}. Magnetic reconnection could also be a possible mechanism to convert the internal energy into the prompt emission \citep{2015AdAst2015E..22P}. Later on, this material expanse and interact with the pre-existing material surrounding the GRB events (external shock), producing the broadband synchrotron radiation, which attributes to the afterglow emission \citep{2004RvMP...76.1143P, 2010ApJ...720.1513K}. The external shock consists of two shock waves, the forward shock (FS) moving towards the external medium accompanied by a reverse shock (RS) which will travel into the ejecta itself. Generally, RS emission described the early afterglow data, seen in a few GRBs, and most of the afterglows data are well explained using the FS model. FS emission helps constrain the burst's circumburst medium, jet geometry, and total energy \citep{2004RvMP...76.1143P}. Furthermore, GRBs are also categorized based on the absence of emission in a particular wavelength; for example, the lack of optical emission leads to ``Dark bursts," and absence of gamma-ray emission leads to ``orphan GRBs."

{\bf Dark GRBs:} The first GRB afterglow (in X-ray) associated with GRB 970228 was discovered in 1997 by BeppoSAX mission \citep{1997Natur.387..783C}. Later on, an optical afterglow was also detected from ground-based follow-up observations for the same burst at redshift $z$ = 0.695 \citep{1997Natur.386..686V}. However, soon after the first discovery of the optical afterglow, no optical counterpart associated with GRB 970828 was detected despite deep searches \citep{1998ApJ...493L..27G}, and the number of such GRBs (with an X-ray counterpart, but no optical counterpart) is increasing. These bursts are defined as ‘dark burst’ or ‘optically dim burst’ \citep{2004ApJ...617L..21J, 2003A&A...408L..21P}. In the first instance, the non-detection of optical counterparts was explained due to the delayed follow-up observations (counterpart had faded significantly below the telescopes sensitivity limit) due to the unavailability of precise localization. However, after the launch of \swift mission in 2004, rapid follow-up observations of afterglows helped reduce the fraction of dark GRBs, but still a significant fraction of dark bursts exits. From 1997 to 2020, $\sim$ 66 \% (1443/2173) of well-localized GRBs are detected with an X-ray counterpart by various X-ray missions; however, so far, only $\sim$ 38 \% (831/2173) GRBs are detected with an optical counterpart\footnote{\url{https://www.mpe.mpg.de/~jcg/grbgen.html}}.

In the present era of \swift mission, dark GRBs have been define in the framework of the most accepted fireball model of GRBs. \cite{2003ApJ...592.1018D} propose to define the dark burst using the ratio of optical to X-ray flux. \cite{2004ApJ...617L..21J} propose to use the optical to X-ray spectral index ($\beta_{\rm OX}$) to define the dark GRBs ($\beta_{\rm OX} < 0.5$). \cite{2009ApJ...699.1087V} propose to use the optical ($\beta_{\rm O}$) and X-ray ($\beta_{\rm X}$) spectral indices depending on the spectral regime. For example, $\beta_{\rm X}$ = $\beta_{\rm O}$ + 0.5, if the cooling frequency ($\nu_{c}$) of synchrotron spectrum is located in between the X-ray and optical frequencies and $\beta_{\rm X}$ = $\beta_{\rm O}$, for all other possible spectral regime. Therefore, the possible range of $\beta_{\rm OX}$ in all the possible spectral regime is $\beta_{\rm X}$ -0.5 $\leq$ $\beta_{\rm OX}$ $\leq$ $\beta_{\rm X}$. In this context, they classified the dark GRBs by $\beta_{\rm OX} < \beta_{\rm X}-0.5$.  

There are various possible different factors responsible for the optical darkness of afterglows \citep{2011A&A...526A..30G}. i) optical afterglows could be intrinsically faint; ii) GRBs could be detected at a high redshift, because of which the Lyman-$\alpha$ forest emission will affect the optical emission; iii) Obscuration scenario, this could be due to dust in the GRBs host galaxies at larger distances or along the line of sight so that this could cause for a very reddened optical afterglow.

{\bf Orphan Afterglows:} In the present era of GRBs, many space-based missions such as \swift, \fermi, \kw, \AstroSat, {\it INTEGRAL}, etc., are continuously searching the whole sky for new GRBs candidates (prompt emission) with a large field of view (FOV). However, suppose a burst has a viewing angle ($\theta_{obs}$) greater than the jet opening angle ($\theta_{core}$). In that case, i.e., the case of off-axis observations, no gamma-ray emission will be detected as the prompt emission is beamed within an angle $1/\Gamma_{0} <$ $\theta_{core}$, where $\Gamma_{0}$ is the bulk Lorentz factor, \citep{2002ApJ...576..120T, 2014PASA...31...22G}. Therefore, space-based missions can only discover those GRBs whose jet is directed towards the Earth. But if the beaming angle intercepts the line of sight, multiwavelength afterglow can be detected. Such afterglows without any prompt emission detection are known as ``orphan afterglows." In the current era of survey telescopes having large FOV such as Zwicky Transient Facility (ZTF), and coming facilities like the Large Synoptic Survey Telescope (LSST) help to discover more number of orphan afterglows.

{\bf Afterglow follow-up observations using 3.6m DOT:} Considering India's longitudinal advantage for the follow-up observations of GRBs, deep follow-up observations of possible afterglows of GRBs were occasionally performed \citep{2020GCN.27653....1K, 2020GCN.29148....1P, 2021GCN.29364....1G, 2021GCN.29490....1G, 2021RMxAC..53..113G} using India's largest 3.6-meter Devasthal Optical Telescope and other facilities located at Devasthal observatory of Aryabhatta  Research  Institute of  Observational Sciences (ARIES) Nainital. The optical and near-infrared (NIR) back-end instruments of 3.6m DOT \citep{2020JApA...41...33S, 2018BSRSL..87...42P} offer spectral and imaging capabilities from optical to NIR wavelength and are very important for deep observations of afterglows and other fast fading transients. ARIES has a long history of more than two decays for the afterglow follow-up observations. Deep photometric observations of afterglows are essential for identifying the associated supernovae bumps observed in nearby long bursts, the dark nature of afterglows, jet break, total energy, and their environment. On the other hand, spectroscopic observations are helpful for the redshift measurements of GRBs. In the present work, we have studied a detailed analysis of a potential dark (GRB 210205A) and an orphan afterglow (ZTF21aaeyldq) followed by 3.6m DOT. 
We have organized this paper in the following sections. In sections \ref{GRB 210205A} and \ref{ZTF21aaeyldq}, we present the properties of GRB 210205A and ZTF21aaeyldq, respectively. Finally, in section \ref{Summary and Conclusion}, we have given the summary and conclusion of this work. We have used following cosmological parameters: the Hubble parameter $\rm H_{0}$ = 70 km $\rm s^{-1}$ $\rm Mpc^{-1}$, density parameters  $\rm \Omega_{\Lambda}= 0.73$, and $\rm \Omega_m= 0.27$ \citep{2011ApJS..192...14J}.

\section{GRB 210205A: Dark burst?}
\label{GRB 210205A}

\begin{figure}[!t]
\includegraphics[scale=0.34]{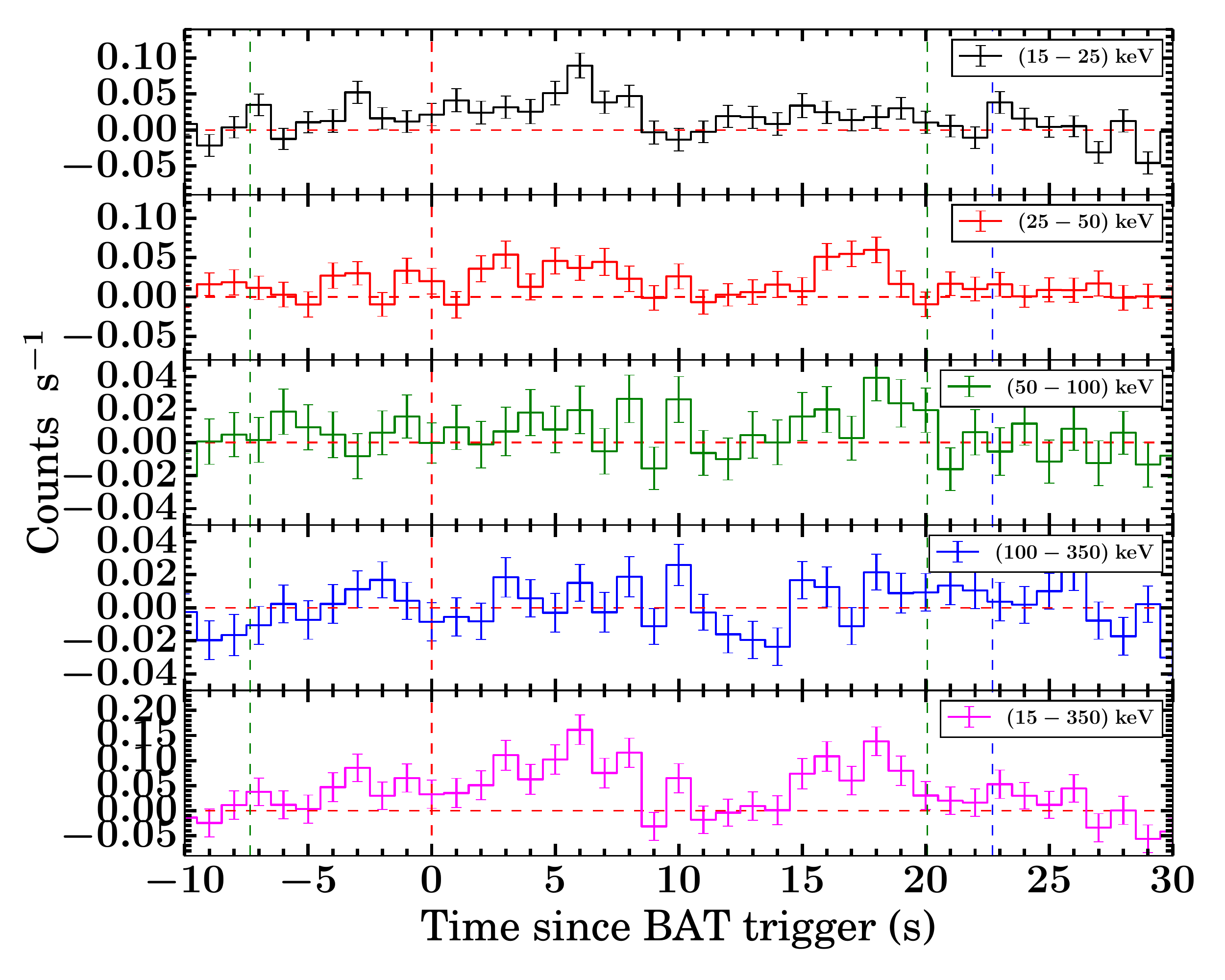}
\caption{The prompt emission light curves of GRB 210205A in different energy channels were obtained using \swift BAT observations. The green vertical dashed lines show the time interval used for time-averaged spectral analysis. The vertical red and blue lines indicate the trigger and end times of \tninty duration, respectively.}\label{batlc}
\end{figure}

\begin{figure}[!t]
\includegraphics[scale=0.34]{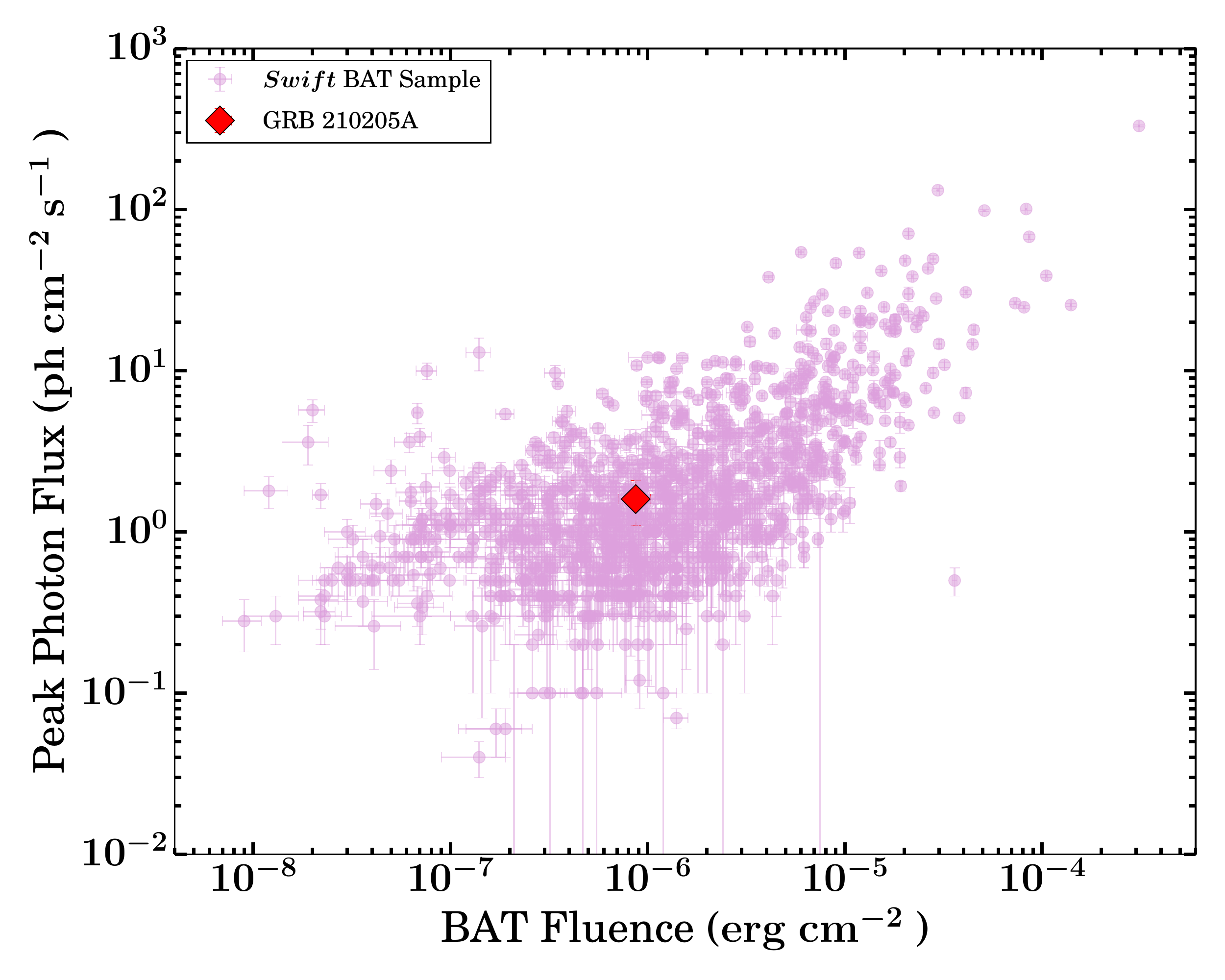}
\caption{The peak photon flux distribution (15-150 \keV) as a function of energy fluence (15-150 \keV) for \swift BAT GRBs. The location of GRB 210205A is shown with a red diamond, showing typical GRB characteristics.}\label{bat_ppf_fluence}
\end{figure}

GRB 210205A was discovered by Burst Alert Telescope (BAT) instrument of NASA's \swift mission at 11:11:17 UT on 05 February 2021 (\swiftT) at the location RA, Dec = 347.257, +56.312 (J2000) with an uncertainty of three arcmin \citep{2021GCN.29397....1D}.
We downloaded and analyzed the BAT data following the method discussed in \cite{2021MNRAS.505.4086G}. The BAT hard X-ray mask-weighted prompt emission light curve comprises of a multi-peaked soft structure with a \tninty duration (in 15-350 \keV)  of 22.70 $\pm$ 4.18 s (see Fig. \ref{batlc}). The time-averaged spectrum from \swiftT -7.35 to \swiftT +20.07 s is best modelled by a simple power-law function with an index of 2.27 $\pm$ 0.31 \citep{2021GCN.29409....1B}. In this temporal window, we calculated the energy fluence equal to (8.7 $\pm$ 1.6) $\rm \times 10^{-7} erg~cm^{-2}$ (in 15-150 \keV). We compared the BAT energy fluence in 15-150 \keV and peak photon flux in the same energy range of GRB 210205A with all the \swift BAT detected GRBs sample. We noticed this GRB is positioned nearly at the center of this distribution (see Fig. \ref{bat_ppf_fluence}), suggesting an intermediate bright GRB.

Due to the narrow spectral coverage of \swift BAT, we calculated the peak energy (\Ep) of the burst using the correlation between observed fluence and peak energy \citep{2020ApJ...902...40Z}, i.e. \Ep = [energy fluence/($\rm 10^{-5} erg~cm^{-2}$)]$^{0.28}$ $\rm \times 117.5^{+44.7}_{-32.4}$ \keV.  We find \Ep equal to 59.31$^{+ 22.56}_{- 16.35}$. The softer value of \Ep further confirms that it was a long-duration GRB. 

\subsection{X-ray afterglow and analysis}

\begin{figure}[!t]
\includegraphics[scale=0.34]{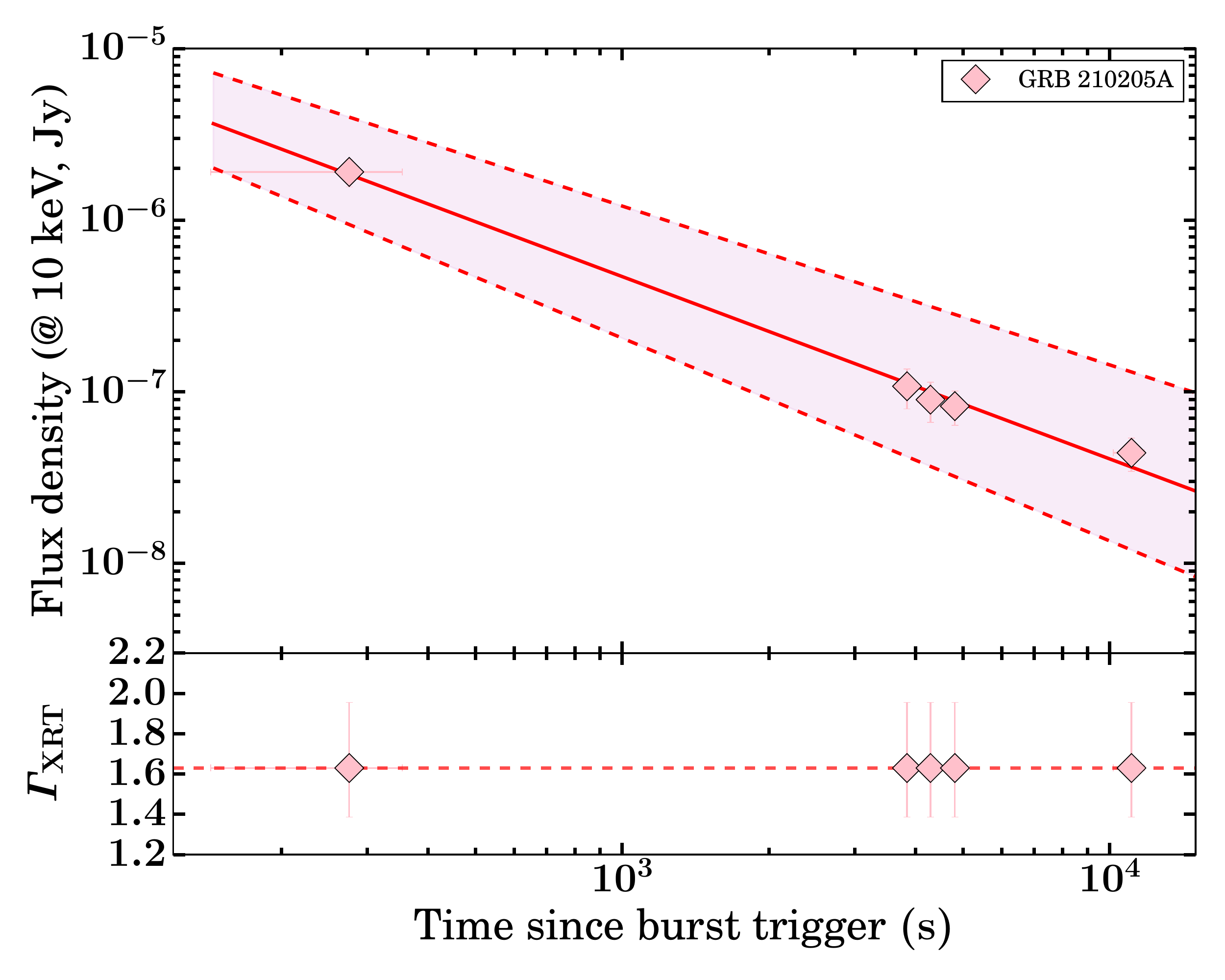}
\caption{{\it Top panel:} Temporal evolution of the X-ray afterglow of GRB 210205A along with a simple power-law model fit. {\it Bottom panel:} Evolution of photon indices in 0.3-10 \keV energy range.}
\label{xrtafterglow_210205A}
\end{figure}

The spacecraft slewed immediately to the burst location to search for the X-ray and optical/UV afterglows. The X-ray telescope (XRT) of \swift detected a new uncatalogued X-ray source (RA, Dec = 347.2214, 56.2943 (J2000)) $\sim$ 134.7 seconds since BAT detection \citep{2021GCN.29397....1D}. For the present work, we retrieved the X-ray data (both light curve and spectrum) products from the \swift XRT online repository \footnote{\url{https://www.swift.ac.uk/}} and followed the analysis methodology discussed in \cite{2021MNRAS.505.4086G}. The X-ray afterglow light curve has been presented in Fig. \ref{xrtafterglow_210205A}. The evolution of X-ray photon indies in 0.3-10 \keV energy range has also been presented in Fig. \ref{xrtafterglow_210205A}. The X-ray afterglow light curve could be best described with a power-law function with a temporal index of 1.06$^{+0.14}_{-0.12}$ (it is in good agreement with the result of the light curve fitting available at the XRT page\footnote{\url{https://www.swift.ac.uk/xrt_live_cat/01030629/}}). 
As no redshift has been reported for this GRB, we modelled the time-averaged X-ray afterglow spectrum (\swiftT + 143 to \swiftT + 39716 s) considering redshift equal to 2, roughly mean redshift value for long bursts. The spectrum could be modelled using an absorption power-law with following spectral parameters: hydrogen column density for the host galaxy ($\rm NH_{\rm host})= 5.77^{+6.26}_{-4.68} \times 10^{22}{\rm cm}^{-2}$ and $\beta_{\rm X}$ = 1.17$^{+ 0.41}_{-0.37}$.
To constrain the spectral regime, we implemented the closure relations for ISM and wind-like medium. We find that X-ray emission is explained with an adiabatic deceleration without an energy injection case. The closure relations also indicate that the X-ray afterglow could be best described with $\nu_{X-ray} > \nu_{c}$ spectral regime for a constant as well as wind ambient medium with the electron energy index p $\sim$ 2.34. 

\begin{figure}[!t]
\includegraphics[scale=0.34]{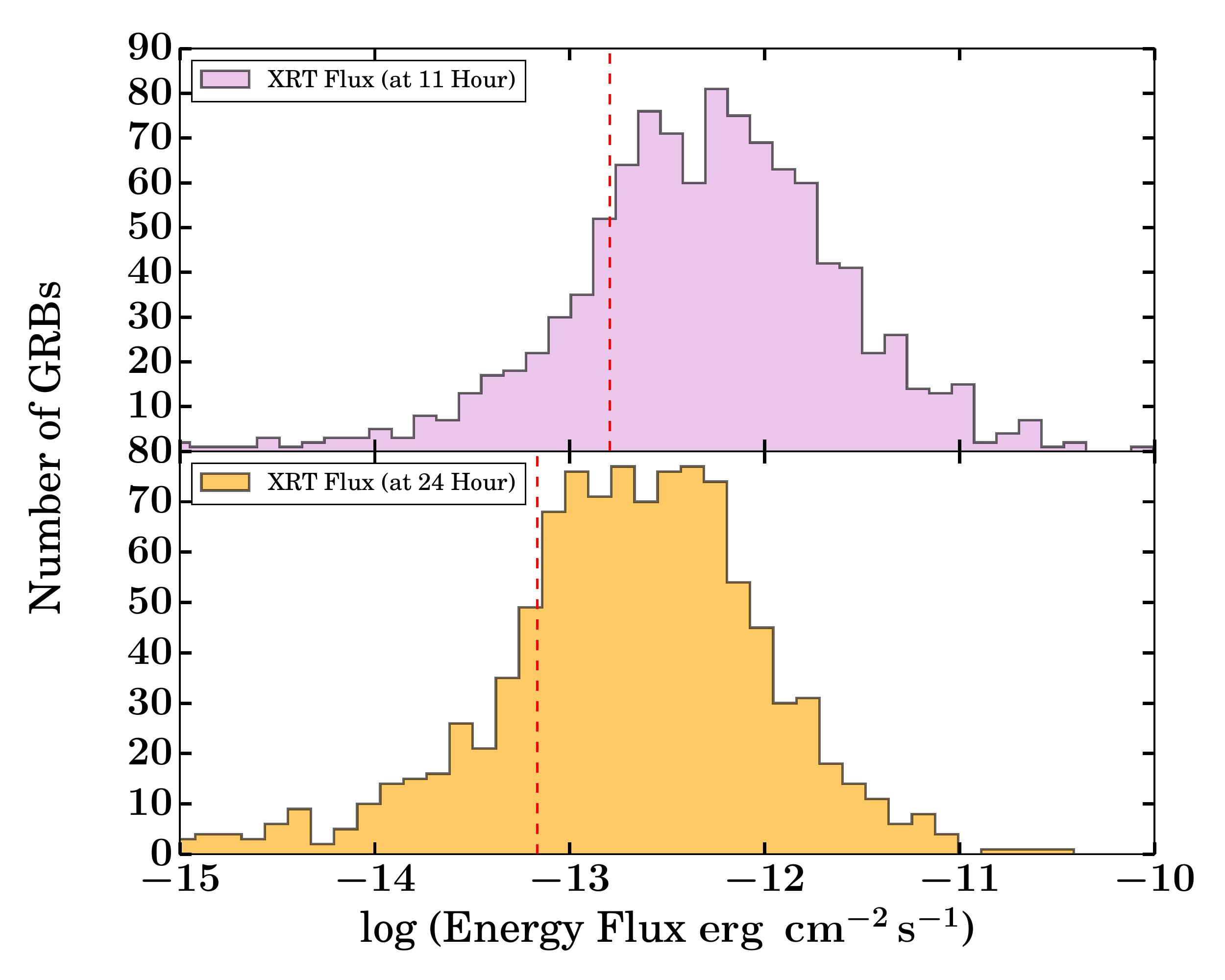}
\caption{{\it Top panel:} The energy flux distribution for a complete sample of X-ray afterglows, detected by \swift XRT at 11 hours after the burst detection in 0.3-10 \keV energy range. {\it Bottom panel:} similar as a top panel but flux calculated at 24-hour post burst. The vertical red dashed lines represent the position of GRB 210205A in respective panels.}
\label{xrtflux}
\end{figure}

Further, we compared the XRT flux of GRB 210205A at 11 hours and 24 hours post burst in 0.3-10 \keV energy range with a complete sample of X-ray afterglows of long GRBs detected by \swift XRT till August 2021. We noticed that the X-ray afterglow of GRB 210205A is faint in comparison to the typical X-ray afterglows at both the epochs (see Fig. \ref{xrtflux}). 

\subsection{Optical follow-up observations and analysis}

\begin{table*}[t]
\caption{The photometric observations log of GRB 210205A taken with 3.6m DOT. The tabulated magnitudes are in the AB magnitude system and have not been corrected for foreground extinction.}
\begin{center}
\begin{tabular}{c c c c c c}
\hline
\bf $\rm \bf T_{mid}$ (days) & \bf Exposure (s)  & \bf Magnitude  &\bf Filter & \bf Telescope & \bf References\\
\hline
1.0921 & 2 x 300 & $> 22.8$ & R & 3.6m DOT & Present work \\
1.1033 & 2 x 300 & $> 22.6$ & I & 3.6m DOT & Present work \\
\hline
\vspace{-2em}
\end{tabular}
\end{center}
\label{tab:observationslog:210205A}
\end{table*}

\swift Ultra-violet and Optical telescope (UVOT) \citep{2021GCN.29397....1D}, and Xinglong GWAC-F60A telescope \citep{2021GCN.29398....1X} started searching for the early optical emission, but no optical afterglow associated with GRB 210205A was detected. We performed the search for any new optical source using 0.6m Burst Observer and Optical Transient Exploring System (BOOTES) robotic telescope $\sim$ 1.18 hours post burst. We did not detect any new source within the \swift XRT enhanced position \citep{2021GCN.29399....1O, 2021GCN.29400....1H}. Furthermore, many other ground-based telescopes also performed deeper observations, but no optical afterglow candidate was reported \citep{2021GCN.29401....1F, 2021GCN.29402....1L, 2021GCN.29406....1H}.

\begin{figure}[ht!]
\includegraphics[angle=0,scale=0.45]{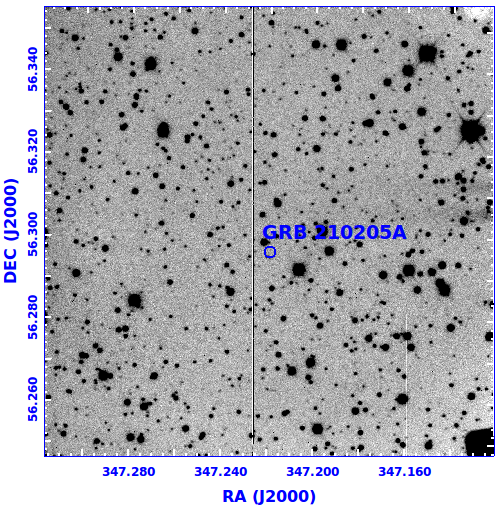}
\caption{The R-band finding chart of GRB 210205A obtained $\sim$ 1.10 days post burst using the 3.6m DOT. The field of view is $\sim$6.5$'$ $\times$ 6.5$'$, and the blue circle indicates the five arcsec uncertainty region at XRT ground localization.}
\label{fig:210205A}
\end{figure}

Further, we started the search for the optical afterglow of this XRT localized burst using the 15 $\mu$m pixel size 4K$\times$4K Charge-coupled device (CCD) Imager placed at the axial port of the newly installed 3.6m DOT of ARIES Nainital. The 4K$\times$4K CCD Imager is capable of deep optical imaging within a field of view of 6.5$'$ $\times$ 6.5$'$ \citep{2018BSRSL..87...42P}. Multiple frames with exposure times of 300 s each were taken in R, and I filters $\sim$ 1.10 days post burst \citep{2021GCN.29526....1P}. We do not find evidence of an optical afterglow source inside the XRT error circle, in agreement with other optical non-detections (see Table \ref{tab:observationslog:210205A}). We constrain the 3-sigma upper limits ($>$ 22.8 mag in R and $>$ 22.6 mag in I filters, respectively). A finding chart obtained using 4K$\times$4K CCD Imager is shown in Fig. \ref{fig:210205A}.

\begin{figure}[ht!]
\includegraphics[angle=0,scale=0.35]{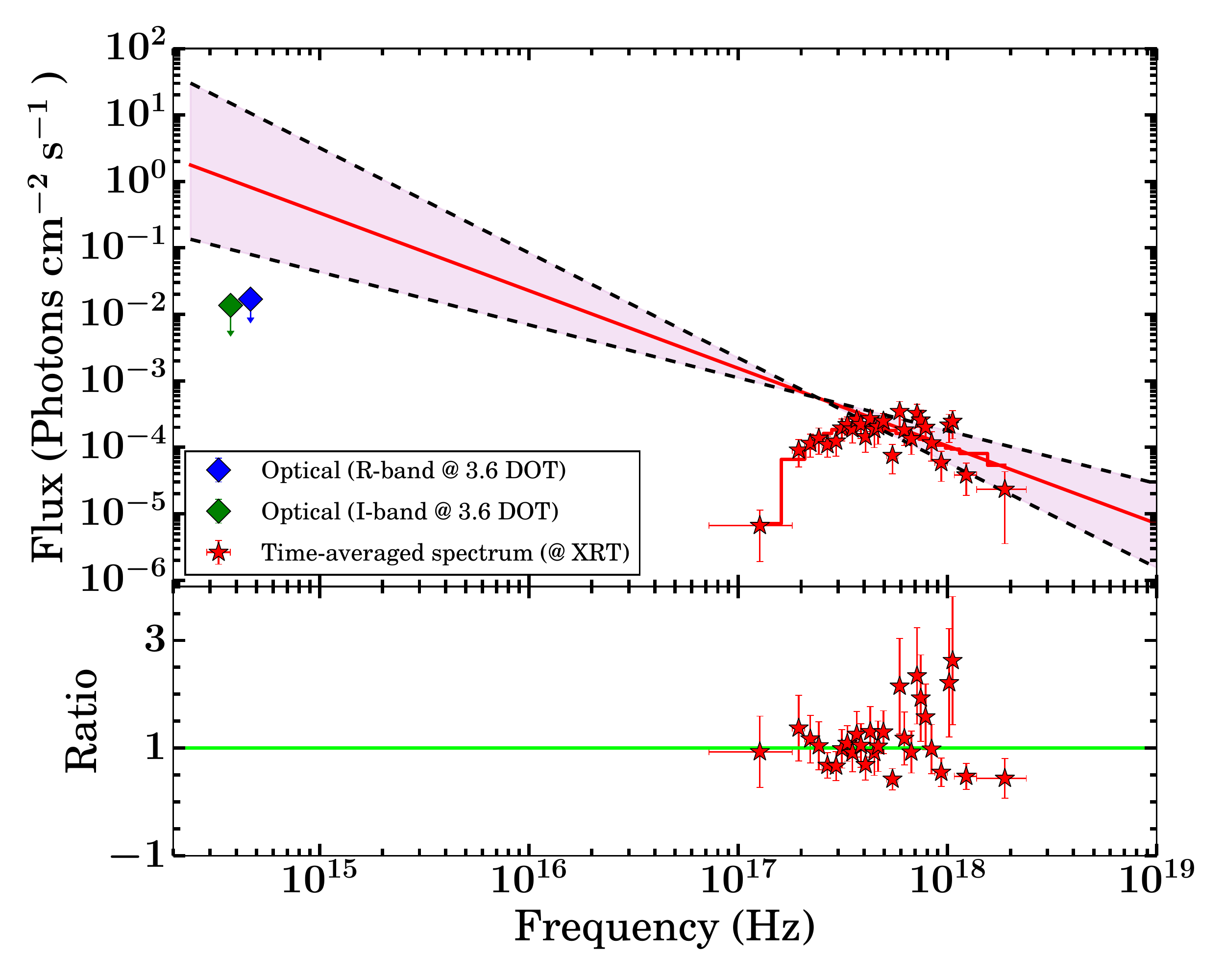}
\caption{{\it Top panel:} The spectral energy distribution for the afterglow of GRB 210205A. The solid red line shows the best fit for the time-averaged XRT spectrum, and the plum shaded color region shows its uncertainty region. Our R and I filters foreground corrected optical data points are shown in blue and green diamonds, respectively. {\it Bottom panel:} Ratio of data and model obtained after the spectral fitting of X-ray data.}
\label{SED:210205A}
\end{figure}

\subsection{Spectral energy distribution}

Spectral energy distribution (SED) is helpful to constrain the afterglow behavior. Considering no spectral break between X-ray and optical frequencies, we extrapolated the X-ray spectral index towards optical frequencies to constrain the upper limit of the intrinsic flux of optical afterglow. We found that our deep limiting magnitude values (Galactic extinction corrected) obtained using 3.6m DOT telescope in R and I filters lie below the extrapolated X-ray power-law slope (see Fig. \ref{SED:210205A}). This suggests that it requires absorption, and GRB 210205A could be a potential dark GRB candidate. We calculated the lower limit of extinction (the host extinction in I filter (A$_{\rm I}$) $>$ 0.25 mag) using the SED.

\begin{figure}[ht!]
\includegraphics[angle=0,scale=0.35]{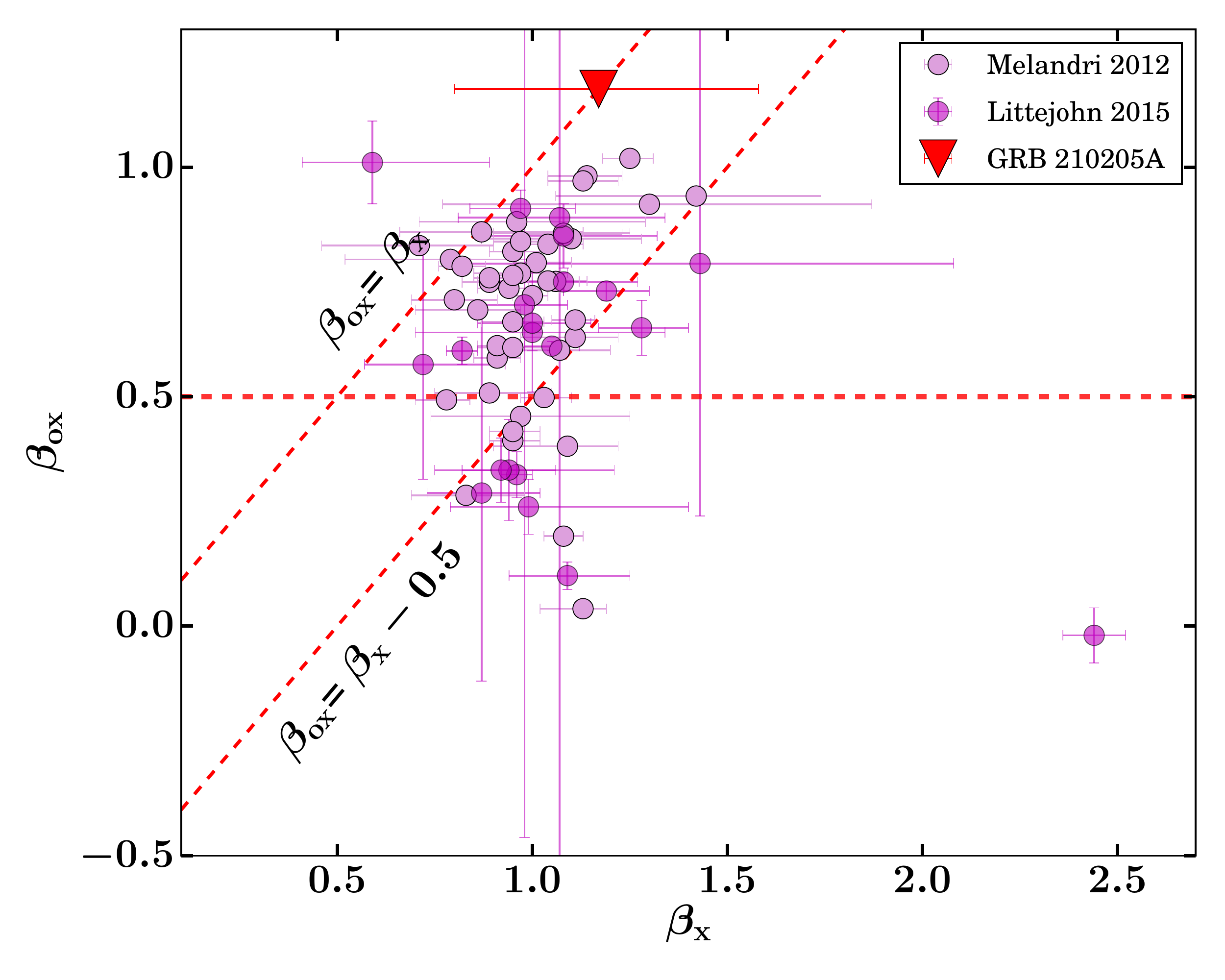}
\caption{Distribution of X-ray spectral indices as a function of optical-X-ray spectral indices for GRB 210205A (shown with a red triangle). For the comparison, data points for bright \swift long GRBs taken from \cite{2012MNRAS.421.1265M, 2015MNRAS.449.2919L} are also shown.}
\label{betaox:210205A}
\end{figure}

Furthermore, we constrain the upper limit on X-ray-to-optical spectral slope ($\beta_{\rm OX}$ $<$ 1.17), as the closure relation suggest for $\nu > \nu_{c}$ spectral regime. We compared the value of $\beta_{\rm OX}$ as a function of $\beta_{\rm X}$ of GRB 210205A along with a large sample of dark population studied by \cite{2012MNRAS.421.1265M, 2015MNRAS.449.2919L}. We find a hint that GRB 210205A satisfies the definition of dark GRBs given by \cite{2009ApJ...699.1087V}. The distribution of $\beta_{\rm OX}$ as a function of $\beta_{\rm X}$ for GRB 210205A is shown in Fig. \ref{betaox:210205A}.

\begin{figure}[ht!]
\includegraphics[angle=0,scale=0.35]{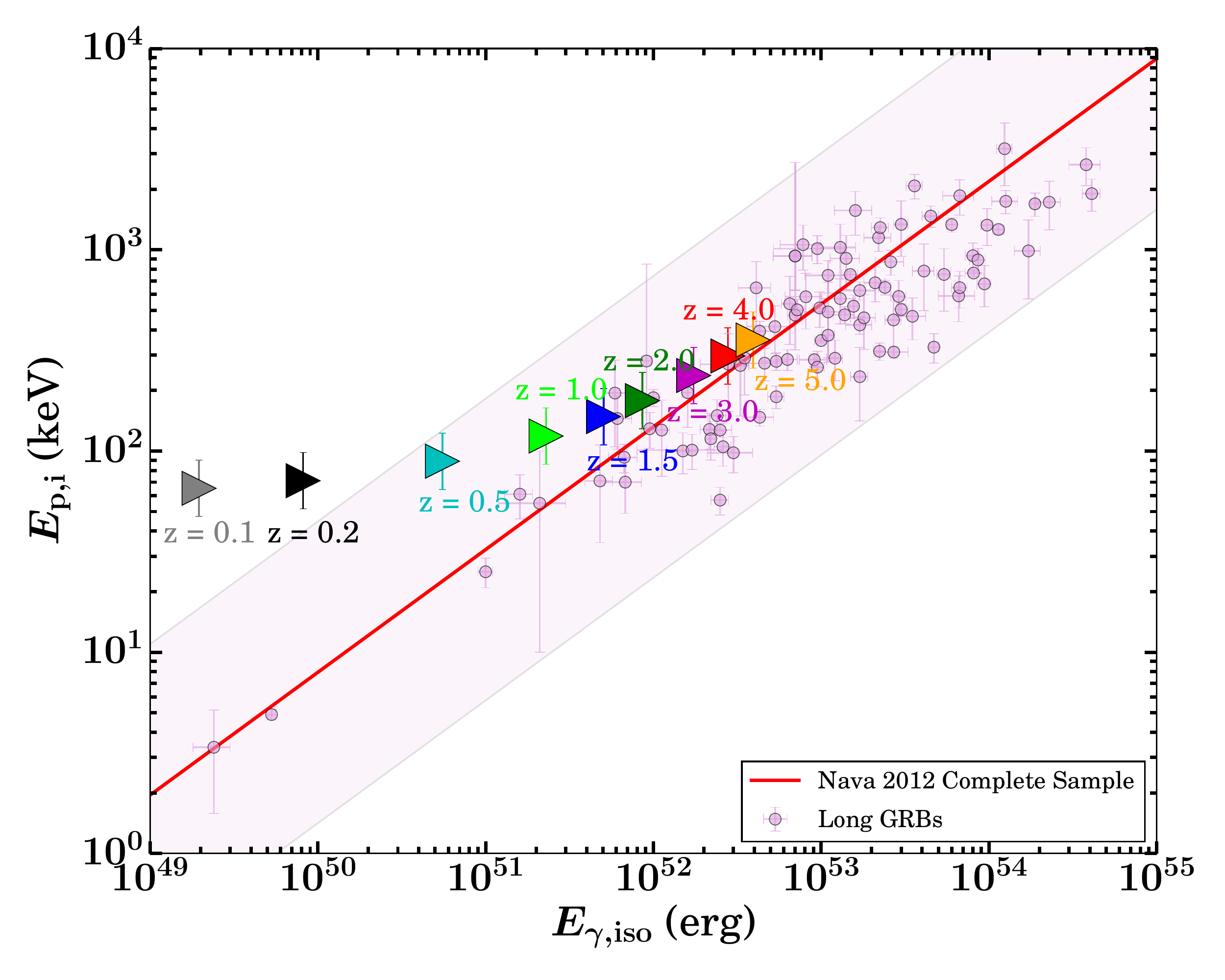}
\caption{GRB 210205A in Amati correlation plane of the long GRBs. The Redshift of the source has been varied from 0.1 to 5 as no spectroscopic or photometric redshift is available. For the comparison, data points for long GRBs taken from \cite{2012MNRAS.421.1256N} are also shown. The solid red line and shaded regions indicate the best fit line and its associated 2-$\sigma$ uncertainty region for the complete sample of long GRBs studied by \cite{2012MNRAS.421.1256N}.}
\label{amati:210205A}
\end{figure}

It is clear that neither optical afterglow is detected nor any host galaxy associated with the burst was reported to measure the redshift of GRB 210205A. So, we have used the Amati correlation to explore the possibility of high redshift origin of the burst. We used BAT fluence value to constrain the lower limit on isotropic gamma-ray energies for a range of redshift values $z$= 0.1 to $z$= 5. Fig. \ref{amati:210205A} shows the position of GRB 210205A for different values of redshift in the Amati correlation plane of long GRBs along with other data points taken from \cite{2012MNRAS.421.1256N}. This analysis indicates that GRB 210205A might be a high redshift burst (with large associated uncertainties). Our analysis suggests that the source was not highly extinguished; therefore, both possibilities, i.e., intrinsically faint or a high redshift origin, could be possible reasons for the optical darkness of this burst.

\section{ZTF21aaeyldq (AT2021any): An orphan afterglow}
\label{ZTF21aaeyldq}

The Zwicky Transient Facility (ZTF) announced the discovery of ZTF21aaeyldq (AT2021any) at the location RA= 08:15:15.34, DEC= -05:52:01.2 (J2000) with $r$ filter magnitude of 17.90 (AB). The object was discovered at 06:59:45.6 UT on 16$^{th}$ January 2021, only 22 minutes after the last non-detection ($r >$  20.28). ZTF carried out two additional observations in the same filter over the next 3.3 hours, and it confirmed that the transient has rapidly faded by two magnitudes \citep{2021GCN.29305....1H}. \cite{2021GCN.29305....1H} suggested that the color of the source is moderately red ($g-r$ $\sim$ 0.3 mag). They also searched for the counterpart/host galaxy, but no such object is visible in deep Legacy Imaging Survey pre-imaging down to a limit of 24 magnitudes \citep{2019AJ....157..168D}. These characteristics suggest that ZTF21aaeyldq is a fast fading, hostless, and Young optical transient. 

\cite{2021GCN.29307....1D} confirm the afterglow behavior of ZTF21aaeyldq using photometric observations and measured the redshift using spectroscopic observations taken with OSIRIS instrument mounted on 10.4m GTC at $\sim$ 16.60 hours post the first detection. They identified many strong absorption lines such as Ly-alpha, SII, OI, SiII, SiIV, CII, CIV, FeII, AlII, and AlIII at a common redshift of $z$ = 2.514 (redshift of ZTF21aaeyldq).

We explored the \swift GRBs archive web page\footnote{\url{https://swift.gsfc.nasa.gov/archive/grb\_table/}} hosted by NASA's Goddard Space Flight Center, \fermi GBM Sub-threshold archive page \footnote{\url{https://gcn.gsfc.nasa.gov/fermi\_gbm\_subthresh\_archive.html}}, \fermi GBM GRBs catalog\footnote{\url{https://heasarc.gsfc.nasa.gov/W3Browse/fermi/fermigbrst.html}}, and GCN Circulars Archive\footnote{\url{https://gcn.gsfc.nasa.gov/gcn3\_archive.html}} for searching the associated gamma-ray counterpart of ZTF21aaeyldq between the last non-detection by ZTF (at 06:39:27 UT on 16 Jan 2021) and the first ZTF detection (at 06:59:46 UT on 16 Jan 2021). No GRB associated with ZTF21aaeyldq is detected using any space-based $\gamma$-ray telescopes during this temporal window. However, on the same day ($\sim$ 46 minutes before the last non-detection of ZTF) \AstroSat Cadmium Zinc Telluride Imager (CZTI) and Large Area X-ray Proportional Counter (LAXPC) detected a burst (GRB 210116A) with a \tninty duration of 9.5$^{+4.1}_{-1.8}$ s \citep{2021GCN.29342....1N} but due to unavailability of precise localization of GRB 210116A, it could not be confirm or rule out the association between both the events. Considering the typical GRBs energy fluence threshold value of $\leq$ $\rm 10^{-6} erg ~{\rm cm}^{-2}$ \citep{2013ApJ...769..130C}, we constrain the isotropic equivalent gamma-ray energy $E_{\rm \gamma, iso}$ $\leq 1.52 \times 10^{52}$ erg.

\begin{figure}
\includegraphics[angle=0,scale=0.47]{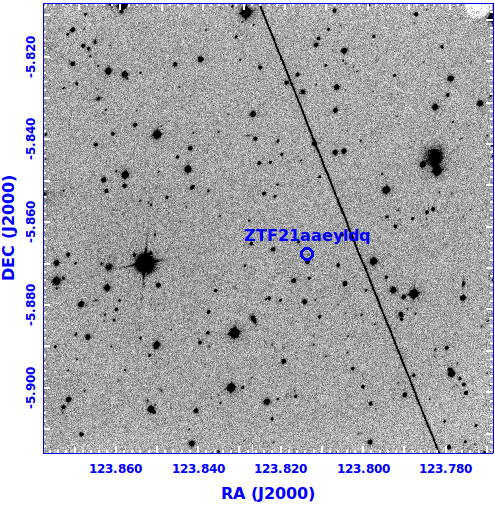}
\caption{The R-band finding chart of ZTF21aaeyldq obtained using the 3.6m DOT (taken $\sim$ 0.65 days post ZTF last non-detection) is shown. The field of view is $\sim$6.5$'$ $\times$ 6.5$'$, and the blue circle indicates the five arcsec uncertainty region at ZTF ground localization.}
\label{fig:ZTF}
\end{figure}

We observed ZTF21aaeyldq using the 4K$\times$4K CCD Imager installed at the axial port of the 3.6m DOT, started $\sim$ 15.258 hours after the ZTF first detection \citep{2021GCN.29308....1K}. We obtained two images with an exposure time of 300 seconds each in Bessel R and I filters. We reduced the data following the method discussed in \cite{2021MNRAS.505.4086G}. We also detected an uncatalogued source in both R, and I filters at the location reported by \cite{2021GCN.29305....1H}. A finding chart (R-filter) showing the detection of ZTF21aaeyldq, obtained using the 4K$\times$4K CCD Imager mounted at the 3.6m DOT, is presented in Fig. \ref{fig:ZTF}. Further, we again observed the source in the r filter at $\sim$ 9.40 days after the first discovery using the same instrument and telescope. We acquired a consecutive set of 12 images with an exposure time of 300 seconds each. We do not detect any optical counterpart up to a magnitude limit of 23.98 in the stacked image \citep{2021GCN.29364....1G}.

In addition to follow-up of ZTF21aaeyldq using 3.6m DOT, we also performed the observations of this source using the 2.2m CAHA telescope located in Almeria (Spain) equipped with the CAFOS instrument. We obtained multiple images in BVRI filters with an exposure time of 240 s for each frame, starting at 00:02:24 UT on 17$^{th}$ January 2021. The optical counterpart is clearly detected in R and I filters and marginally visible in the V filter at the ZTF location. Using this telescope, we again monitored the object at the second epoch in the same filter system BVRI, starting at 04:08 UT. However, the source could not be detected during our second epoch of observations. We reduced the data using the same methodology as we did for 3.6m DOT. We calibrated the instrumental magnitudes with the nearby stars present in the USNO-B1.0 catalog (same stars used for 3.6m DOT data calibration). We have listed the optical photometry log of our observations of ZTF21aaeyldq along with those obtained using different gamma-ray coordination networks (GCNs) in Table \ref{tab:observationslog}.

\begin{table*}[t]
\caption{The optical photometric observations log of ZTF21aaeyldq taken with 3.6m DOT and 2.2m CAHA, including data from GCNs. The tabulated magnitudes are in the AB magnitude system and have not been corrected for foreground extinction. All the upper limits are given with three sigma.}
\begin{center}
\begin{tabular}{c c c c c c}
\hline
\bf $\rm \bf T_{mid}$ (days) & \bf Exposure (s)  & \bf Magnitude  &\bf Filter & \bf Telescope & \bf References\\
\hline
 0.0141 & -- & 17.90 $\pm$ 0.06 & r & ZTF & \cite{2021GCN.29305....1H} \\
 0.6998 & 1 x 60 & 21.64 $\pm$ 0.03  & r & 10.4m GTC  & \cite{2021GCN.29307....1D} \\
0.8451 & 5 x 300 &  21.86 $\pm$ 0.04  & r & NOT &  \cite{2021GCN.29310....1Z} \\
0.8261 & 5 x 300  &  22.28 $\pm$ 0.05  & g  & NOT  & \cite{2021GCN.29310....1Z} \\
1.8230 & --  & 23.39 $\pm$ 0.11 & g & 2.2m MPG &  \cite{2021GCN.29330....1G} \\
 2.0199 & --  &  23.47 $\pm$  0.06 & g & 2.2m MPG & \cite{2021GCN.29330....1G} \\
 2.8700 & --  &  23.84 $\pm$ 0.09  & g & 2.2m MPG & \cite{2021GCN.29330....1G} \\
0.8641 & 5 x 300 & 21.65 $\pm$ 0.03 & i & NOT &  \cite{2021GCN.29310....1Z}  \\
1.0556 & 3 x 180  & 22.10 $\pm$ 0.20  & i & Lowell  & \cite{2021GCN.29309....1A}\\
0.9969 & 1 x 900 & 21.62 $\pm$ 0.20  & J & LBT  &  \cite{2021GCN.29327....1R}\\
0.9969 & 1 x 900 & 21.36 $\pm$  0.24 & H & LBT  &  \cite{2021GCN.29327....1R}\\
1.7584 & 10 x 360 & 22.89 $\pm$ 0.12 & Rc & 2.2m CAHA  & \cite{2021GCN.29321....1K} \\
\hline 
0.7259 & 1 x 240 & $>$ 22.97 & B & 2.2m CAHA  & Present work \\
0.8964 & 1 x 240 & $>$ 22.13 & B & 2.2m CAHA  & Present work \\
0.7297 & 1 x 240 & 22.17 $\pm$ 0.39 & V & 2.2m CAHA  & Present work \\
0.8998 & 1 x 240 & $>$ 21.40 & V & 2.2m CAHA  & Present work \\
0.7332 & 1 x 240 & 21.22 $\pm$ 0.12 & R & 2.2m CAHA  & Present work \\
0.9032 & 1 x 240 & $>$ 20.96 & R & 2.2m CAHA  & Present work \\
0.7367 & 1 x 240 & 21.23 $\pm$ 0.19 & I & 2.2m CAHA  & Present work \\
0.9066 & 1 x 240 & $>$ 20.01 & I & 2.2m CAHA  & Present work \\

0.6498 & 1 x 300 & 21.25 $\pm$ 0.06 & R & 3.6m DOT & Present work\\

0.6546 & 1 x 300 & 21.36 $\pm$ 0.08 & I & 3.6m DOT & Present work\\
9.3963 & 12 x 300 & $> 23.98 $ & r & 3.6m DOT & Present work\\
\hline
\vspace{-2em}
\end{tabular}
\end{center}
\label{tab:observationslog}
\end{table*}

\subsection{Afterglow behaviour of ZTF21aaeyldq}

\begin{table*}[t]
\caption{The observations log for the X-ray afterglow of ZTF21aaeyldq taken with \swift XRT.}
\begin{center}
\begin{tabular}{c c c c c c}
\hline
\bf Obs Ids & \bf $\rm \bf T_{mid}$ (days) &  \bf Count Rate  &\bf Energy range  & \bf Flux & \bf Telescope\\
\hline
00013991001 & 1.02 & 6.66$^{+1.84}_{-1.84}$  & 0.3-10 \keV & 2.99 $\rm \times 10^{-13} erg ~{\rm cm}^{-2} ~{\rm s}^{-1}$ & \swift XRT \\
00013991002 & 4.24 & $<$ 0.0043513 & 0.3-10 \keV & $<$ 1.95 $\rm \times 10^{-13} erg ~{\rm cm}^{-2} ~{\rm s}^{-1}$ & \swift XRT \\
00013991003 & 9.35 & $<$ 0.0040133 & 0.3-10 \keV & $<$ 1.80 $\rm \times 10^{-13} erg ~{\rm cm}^{-2} ~{\rm s}^{-1}$ & \swift XRT \\
\hline
\vspace{-2em}
\end{tabular}
\end{center}
\label{tab:ZTFXRTobservationslog}
\end{table*}

After the independent discovery of the afterglow candidate by ZTF, several ground-based telescopes detected the source in different filters\footnote{\url{https://gcn.gsfc.nasa.gov/other/ZTF21aaeyldq.gcn3}}. In addition to multi-band optical/NIR observations, \cite{2021GCN.29313....1H} reported the detection of X-ray afterglow based on a target-of-opportunity (ToO) observations obtained using \swift XRT, $\sim$ 19.82 hours after the last ZTF non-detection. We obtained and reduced the XRT data using the online tool known as Build \swift-XRT products\footnote{\url{https://www.swift.ac.uk/user$_$objects/}} provided by the \swift team. \swift-XRT observed the source at three different epochs (Obs Ids: 00013991001, 00013991002, and 00013991003) with a total exposure time of 8.2 ks (see Table \ref{tab:ZTFXRTobservationslog}). However, the source is only detected at the first epoch ($\rm \Delta T_{mid}$ $\sim$ 24.50 hours post ZTF last non-detection) at the location of ZTF21aaeyldq with a count rate of 6.66$^{+1.84}_{-1.84} ~\times 10^{-3}$  s$^{-1}$ in 0.3-10 \keV. 
Considering the Galactic hydrogen column density value $\rm NH_{\rm Gal}= 7.75 \times 10^{20}{\rm cm}^{-2}$ \citep{2013MNRAS.431..394W} and X-ray photon index value of 2, we calculated the unabsorbed flux equal to 2.99 $\rm \times 10^{-13} erg ~{\rm cm}^{-2} ~{\rm s}^{-1}$ and X-ray luminosity equal to 1.60 $\rm \times 10^{46} erg ~{\rm s}^{-1}$ at $z$ equal to 2.514. The measured X-ray luminosity is typical of GRB X-ray afterglows at this epoch. We have used Portable, Interactive Multi-Mission Simulator (\sw{PIMMS}) tool of NASA's HEASARC page\footnote{\url{https://heasarc.gsfc.nasa.gov/cgi-bin/Tools/w3pimms/w3pimms.pl}}to calculate the unabsorbed flux. During last two epochs observations, no X-ray emission has been detected at the ZTF afterglow positions. We obtained 3-$\sigma$ upper limits on count rate equal to 0.0043513 ($\rm \Delta T_{mid}$ $\sim$ 4.24 days post ZTF last non-detection) and 0.0040133 ($\rm \Delta T_{mid}$ $\sim$ 9.35 days post ZTF last non-detection) at second and third epoch of observations, respectively. Considering the same values of $\rm NH_{\rm Gal}$ and X-ray photon index, we converted these count rates into an upper limit on the flux density of $<$ 1.95 $\rm \times 10^{-13} erg ~{\rm cm}^{-2} ~{\rm s}^{-1}$ and $<$ 1.80 $\rm \times 10^{-13} erg ~{\rm cm}^{-2} ~{\rm s}^{-1}$, respectively.

\begin{figure}[ht!]
\includegraphics[angle=0,scale=0.35]{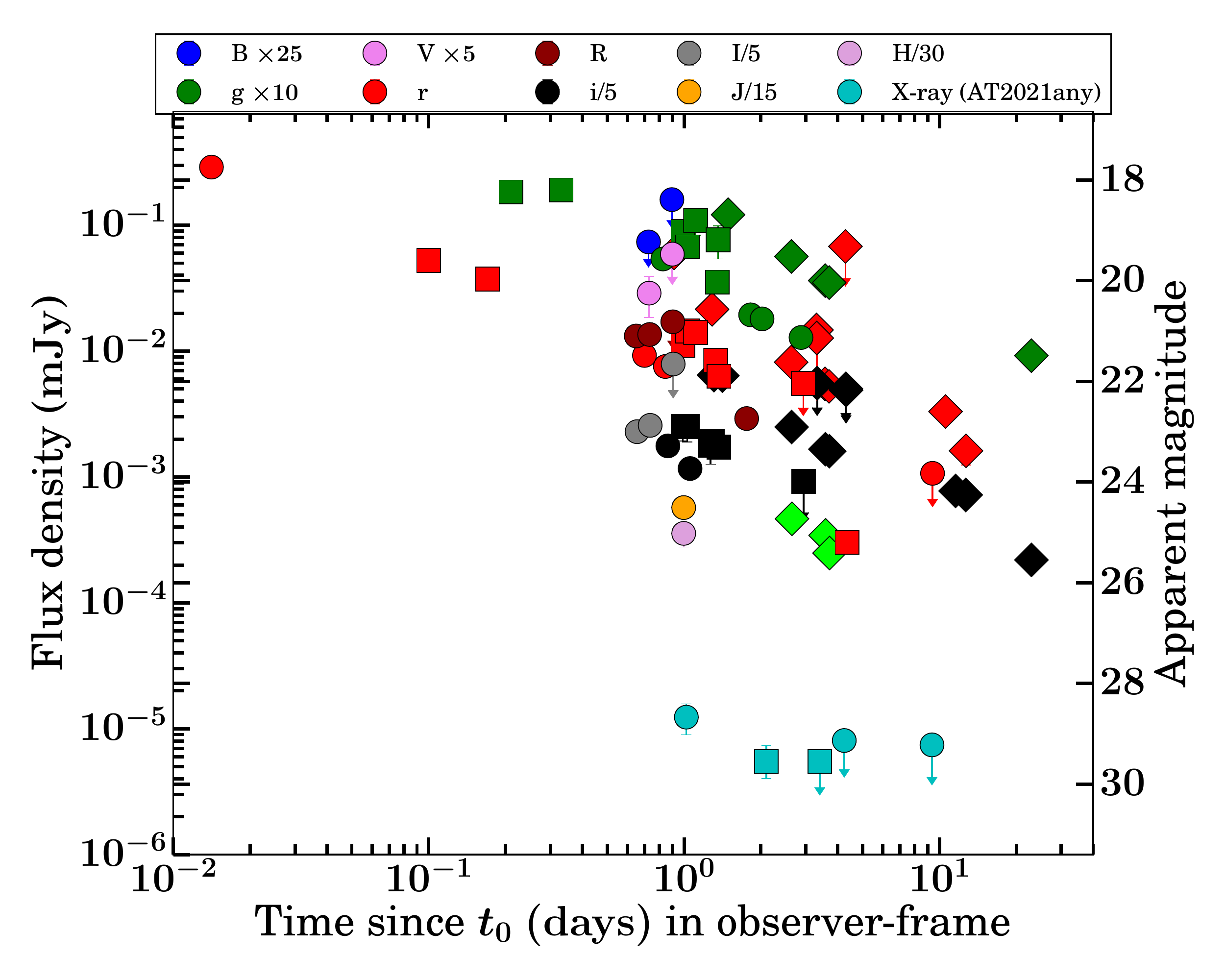}
\caption{{\bf Comparison between afterglow light curves of orphan afterglows with a measured redshift known so far:} The X-ray and optical/NIR afterglow light curve of ZTF21aaeyldq (shown with circles of different colors). For the comparison, we have also demonstrated X-ray and optical afterglow light curves of ZTF20aajnksq ($z \sim$ 2.90), shown with squares and ZTF19abvizsw ($z \sim$ 1.26), pictured with diamonds. We have used the same colors and offsets for ZTF20aajnksq and ZTF19abvizsw also. The lime diamonds denote the z filter observations (with an offset factor of 35) of ZTF19abvizsw.}
\label{lightcurvecomparision}
\end{figure}

The afterglow light curve of ZTF21aaeyldq has been shown in Fig. \ref{lightcurvecomparision}. The photometric magnitudes have been corrected for foreground extinction before converting them into flux units. The optical afterglow light curve of ZTF21aaeyldq has been continuously fading since the first detection by ZTF, typical characteristics of afterglows. We fitted the early $r$ band data taken from \cite{2021GCN.29305....1H, 2021GCN.29307....1D, 2021GCN.29310....1Z} and found that the $r$ band light curve is best described with a single power-law model with a temporal decay slope $\alpha$ equal 0.89 $\pm$ 0.03, consistent with the slope reported by \cite{2021GCN.29321....1K}. Later on, \cite{2021GCN.29344....1K} observed the afterglow candidate ZTF21aaeyldq with the 2.2m telescope at Calar Alto, Spain, in the Rc filter at 2.8 days after the first detection, and with Gamma-ray Burst Optical/Near-infrared Detector (GROND) mounted at the 2.2m MPG telescope at 3.9 days after the first ZTF detection, respectively. They detected the source clearly in each stacked frame. Further, they fitted the optical data taken from various GCNs \citep{2021GCN.29305....1H, 2021GCN.29307....1D, 2021GCN.29310....1Z, 2021GCN.29330....1G} along with their observations. They found that a broken power-law is better fitting the data ($\chi^{2}$/dof = 0.12) with temporal index before the break ($\alpha_{1}$) equal to 0.95 $\pm$ 0.03, temporal index after the break ($\alpha_{2}$) equal to 2.30 $\pm$ 0.76, and break time T$_{b}$ equal to 0.82 $\pm$ 0.08 days. Our late-time observations using 3.6m DOT ($\sim$ 9.40 days post first detection) are also consistent with the temporal index after the break suggested by \citep{2021GCN.29344....1K}. However, we could not confirm if the break is achromatic/chromatic due to the unavailability of simultaneous multi-wavelength data. In any case, the jet break is a typical characteristic of GRB afterglows, and it indicates that the nature of ZTF21aaeyldq is a GRB afterglow. Since this event was established as an orphan afterglow based on no detection of any associated GRBs between the last non-detection and the first detection of optical emission by ZTF, we extended our analysis by comparing the properties with other orphan events with a redshift measurement. The light curve evolution of ZTF21aaeyldq with other known orphan afterglows (ZTF20aajnksq/AT2020blt and ZTF19abvizsw/AT2019pim) are shown in Fig. \ref{lightcurvecomparision}. In the case of ZTF20aajnksq, we collected data from \cite{2020ApJ...905...98H} and for ZTF19abvizsw ($\rm T_{0}$: at 07:35:02 UT on 1$^{st}$ September 2019, the last non-detection\footnote{\url{https://www.wis-tns.org/object/2019pim}}), we obtained optical observations from \cite{2020ApJ...905..145K}. Based on the above, we noticed that ZTF21aaeyldq has many similar features with classical GRBs afterglows. The measured redshift of ZTF21aaeyldq is typical of long GRBs, and the strong absorption lines identified in the optical spectrum are usually present in GRBs afterglows at measured redshift. The detection of X-ray counterpart with typical X-ray luminosity also indicates the afterglow nature of ZTF21aaeyldq. In addition to these, the presence of a break in the optical light curve with typically expected temporal indices due to jet break confirms the afterglow behavior of the source.

\subsection{Afterglow modelling of ZTF21aaeyldq}

According to the standard external shock fireball model for the afterglows, the X-ray and optical emission from afterglows can be described with synchrotron emission for constant or WIND like external medium \citep{1998ApJ...497L..17S}. The broadband synchrotron spectral energy distribution consists of three break frequencies: the synchrotron cooling frequency $\nu_{\rm c}$, the synchrotron peak frequency $\nu_{\rm m}$, and the synchrotron self-absorption frequency $\nu_{\rm a}$. The self-absorption frequency does not affect the X-ray and optical data at early epochs, and it mainly affects the low-frequency observations of afterglows. Depending on the ordering of these break frequencies, we can constrain the spectral regimes of afterglows at a particular epoch \citep{2003BASI...31...19P}. 

\begin{figure}
\includegraphics[angle=0,scale=0.35]{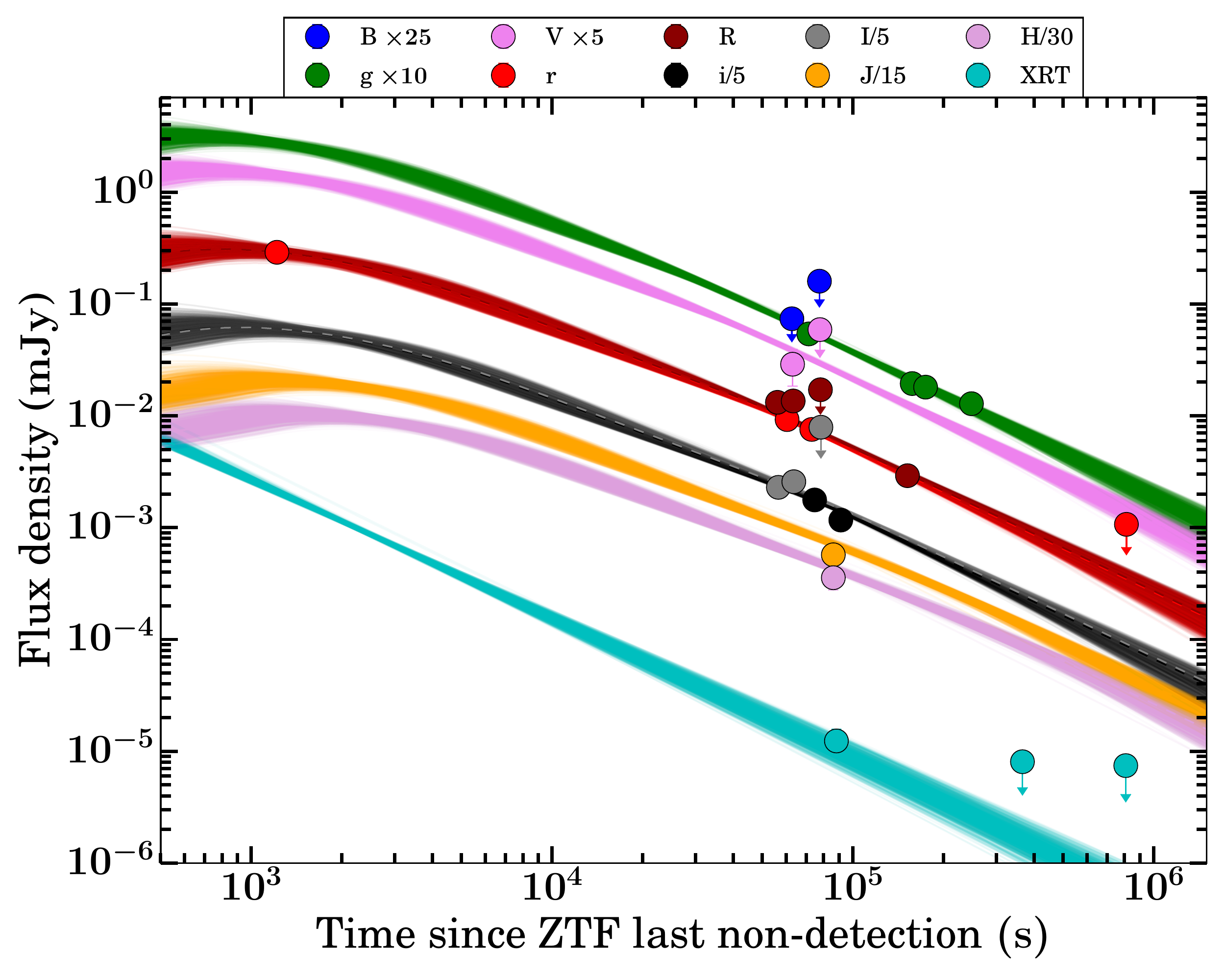}
\caption{Multiwavelength observations data of ZTF21aaeyldq along with the best fit afterglow modelling results obtained using $\af$. The shaded regions indicate the uncertainty region around the median light curve.}
\label{modelling}
\end{figure}

Detailed multiwavelength modelling is helpful to constrain physical parameters associated with the afterglow. We performed detailed multiwavelength modelling of the light curve of the ZTF21aaeyldq afterglow using the publicly available $\af$ Python package. It is an open-source numerical and analytic modelling tool to calculate the multiwavelength light curve and spectrum using synchrotron radiation from an external shock for the afterglows of GRBs \citep{2020ApJ...896..166R}. $\af$ package has capabilities to produce the light curves and spectrum of afterglows considering both structured jets and off-axis observers. We have used Markov chain Monte Carlo (MCMC) Ensemble sampler using \sw{emcee} python package for fitting the multiwavelength light curve to get the model parameters and associated errors.

\begin{figure*}[ht!]
\centering
\includegraphics[angle=0,scale=0.35]{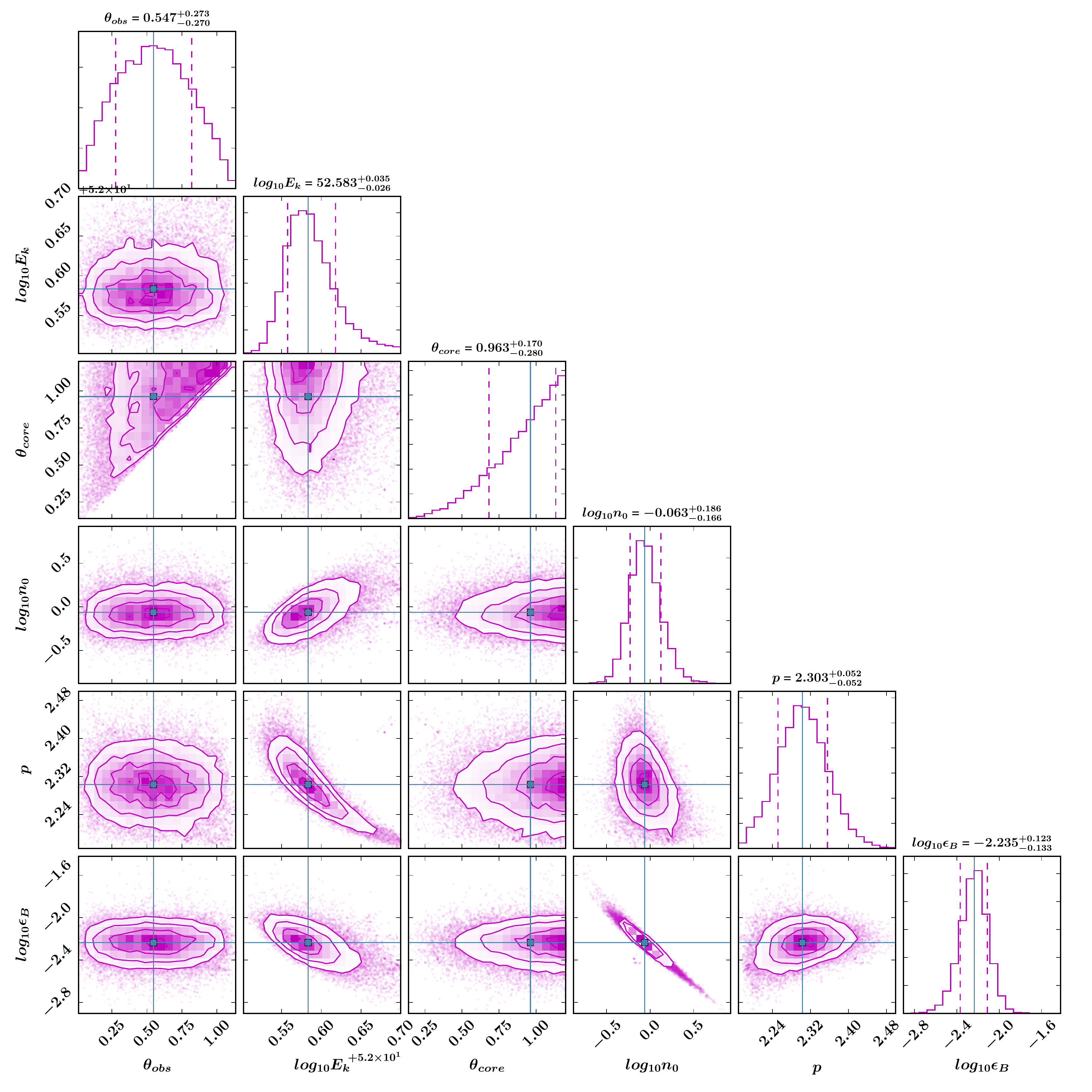}
\caption{Posterior distribution and parameter constraints for the 15000 simulations obtained using broadband afterglow modelling of ZTF21aaeyldq using $\af$.}
\label{modelling_pd}
\end{figure*}

\begin{table*}
\caption{The input parameters, range, best-fit value, and associated errors of multi-wavelength afterglow modelling of ZTF21aaeyldq were performed using the $\af$ python package.}
\begin{center} 
\begin{tabular}{c c c c c}
\hline
Model Parameter & Unit & Prior Type  & Range & Best fit Value \\
\hline \hline
$\theta_{obs}$ & rad & $\sin(\theta_{obs})$ & [0.01, 1.20]  & $0.55_{-0.27}^{+0.27}$ \\
$log_{10}(E_k)$ & erg &  uniform & [51, 52.7] & $52.58_{-0.03}^{+0.03}$\\ 
$\theta_{core}$ & rad & uniform &  [0.01, 1.20] & $0.96_{-0.28}^{+0.17}$  \\ 
$log_{10}(n_0)$ & cm$^{-3}$ &  uniform  &  [-6, 1.6] & $-0.06_{-0.17}^{+0.19}$  \\ 
$p$ & - &  uniform & [2.0001, 2.5] & $2.30_{-0.05}^{+0.05}$   \\ 
$log_{10}(\epsilon_{B})$ & - &  uniform & [-3.6, -1.2] & $-2.23_{-0.13}^{+0.12}$ \\  
$\xi$ & -& -& 1 & - \\
\hline
\hline
\end{tabular}
\end{center}
\label{tab:model}
\end{table*}

There are various jet structures possible for GRBs afterglows. $\af$ has capabilities to produce the light curves for some commonly used jet structures such as top-hat, Gaussian, Power-law, etc. We have used the top-hat jet type structure to model the afterglow data of ZTF21aaeyldq using the $\af$ package. For top hat jet-like afterglows, we consider six free parameters ($\theta_{obs}$: viewing angle, $E_{k}$: on-axis isotropic equivalent energy, $\theta_{core}$: half-width of the jet core, $n_{0}$: number density of ISM medium, $p$: electron distribution power-law index, and $\epsilon_{B}$: thermal energy fraction in magnetic field). We have fixed the $\epsilon_{e}$ (thermal energy fraction in electrons) value = 0.10 \citep{2002ApJ...571..779P}. We have considered sine prior for viewing angle, and for all the remaining parameters, we have used uniform priors. In addition, we assume that the fraction of electrons that get accelerated is equal to one, and the redshift is equal to 2.514 for the model fitting. Furthermore, due to the large scales, we have set $E_{k}$, $n_{0}$ and $\epsilon_{B}$ parameters on the log scale. We have listed the priors types and the parameters bounds for each model parameter in Table \ref{tab:model}.

\begin{figure}[ht!]
\includegraphics[angle=0,scale=0.35]{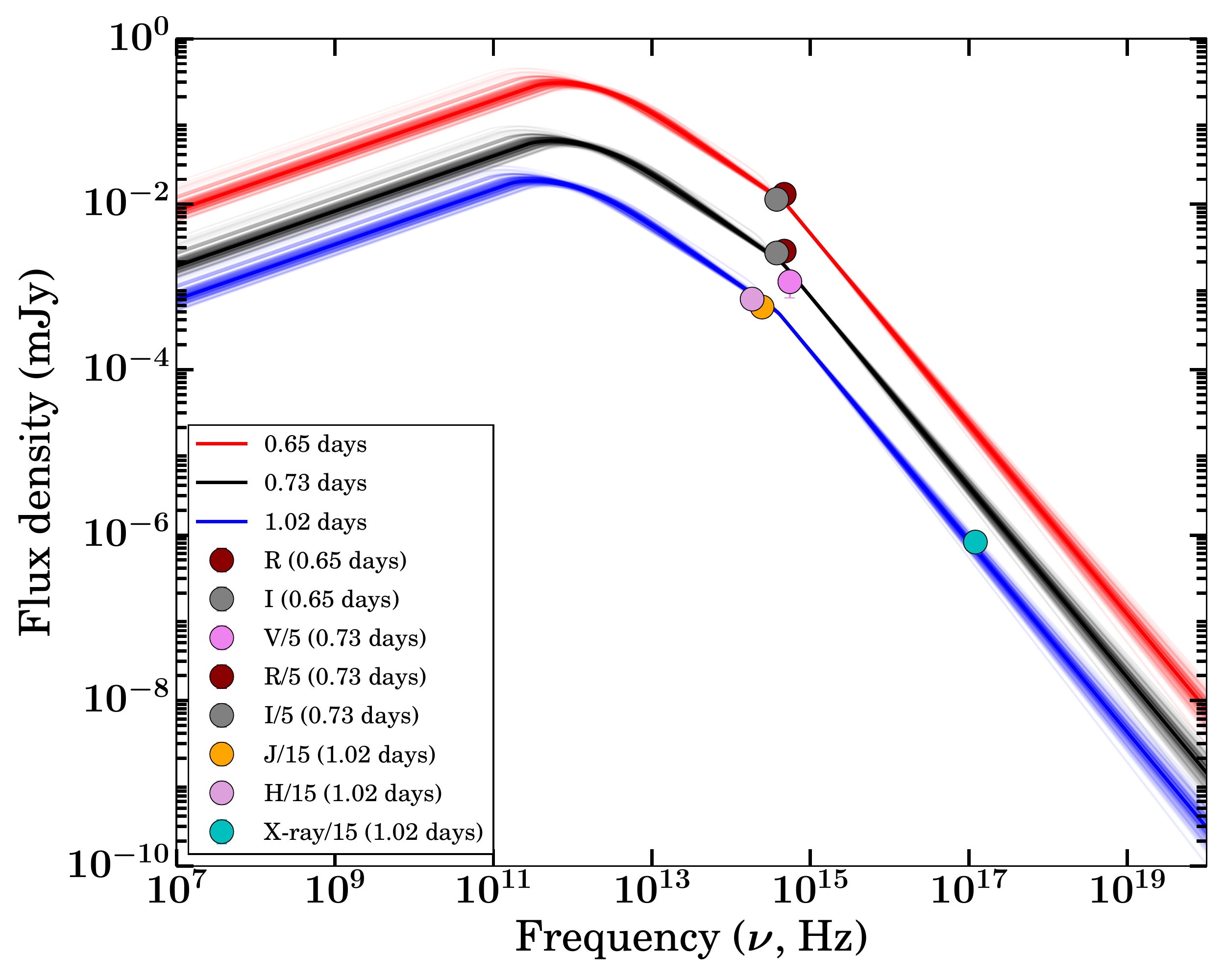}
\caption{The spectral energy distributions of ZTF21aaeyldq at three different epochs obtained for the best fit model parameters using $\af$. The SEDs show the evolution of break frequencies of synchrotron emission.}
\label{modelling_SED}
\end{figure}

In Fig. \ref{modelling}, we have shown the optical and X-ray afterglow light curves of ZTF21aaeyldq along with the best fit model obtained using $\af$.  The observed X-ray and optical data points are shown with circles in the figure. The dark-color dashed lines show the light curve generated using the median value from parameter distribution from the MCMC routine. The shaded colored bands show the uncertainty bands around the median light curve. Fig. \ref{modelling_pd} shows the posterior distribution for the best fit results of microphysical parameters obtained using MCMC simulation. The distribution of median posterior and 16\% and 84\% quantiles for each modeled parameter are also presented in the same plot. We modelled the optical and X-ray afterglow and obtained the following best fit micro-physical parameters: the model constrained the jet on-axis isotropic equivalent energy $log_{10}E_{k}$ = 52.58$^{+0.04}_{-0.03}$ erg, the jet structure parameter $\theta_{core}$ = 0.96$^{+0.17}_{-0.28}$ radian, viewing angle $(\theta_{obs})$ = 0.55$^{+0.27}_{-0.27}$ radian, the ambient number density $log_{10}n_{0}$ = -0.06$^{+0.19}_{-0.17}$ cm$^{-3}$, the parameters electron energy index $p$ = 2.30$^{+0.05}_{-0.05}$, and $log_{10}\epsilon_{B}$ = -2.24$^{+0.12}_{-0.13}$. These parameters are consistent with the typical afterglow parameters of other well studied GRBs \citep{2002ApJ...571..779P}.

Furthermore, we created the SEDs at three epochs (see Fig. \ref{modelling_SED}). The first at 0.65 days (the epoch of 3.6m DOT observations), the second at 0.73 days (the epoch of CAHA observations), and the third at 1.02 days (the epoch of detection of the X-ray emission). During the first and second SEDs, the cooling frequency ($\nu_{c}$) lies close to our optical observations taken with 3.6m DOT and 2.2m CAHA telescope, and later on, it seems that $\nu_{c}$ passes through the optical observations.

\subsection{Possible explanations of the orphan afterglow}

\begin{figure}
\includegraphics[angle=0,scale=0.35]{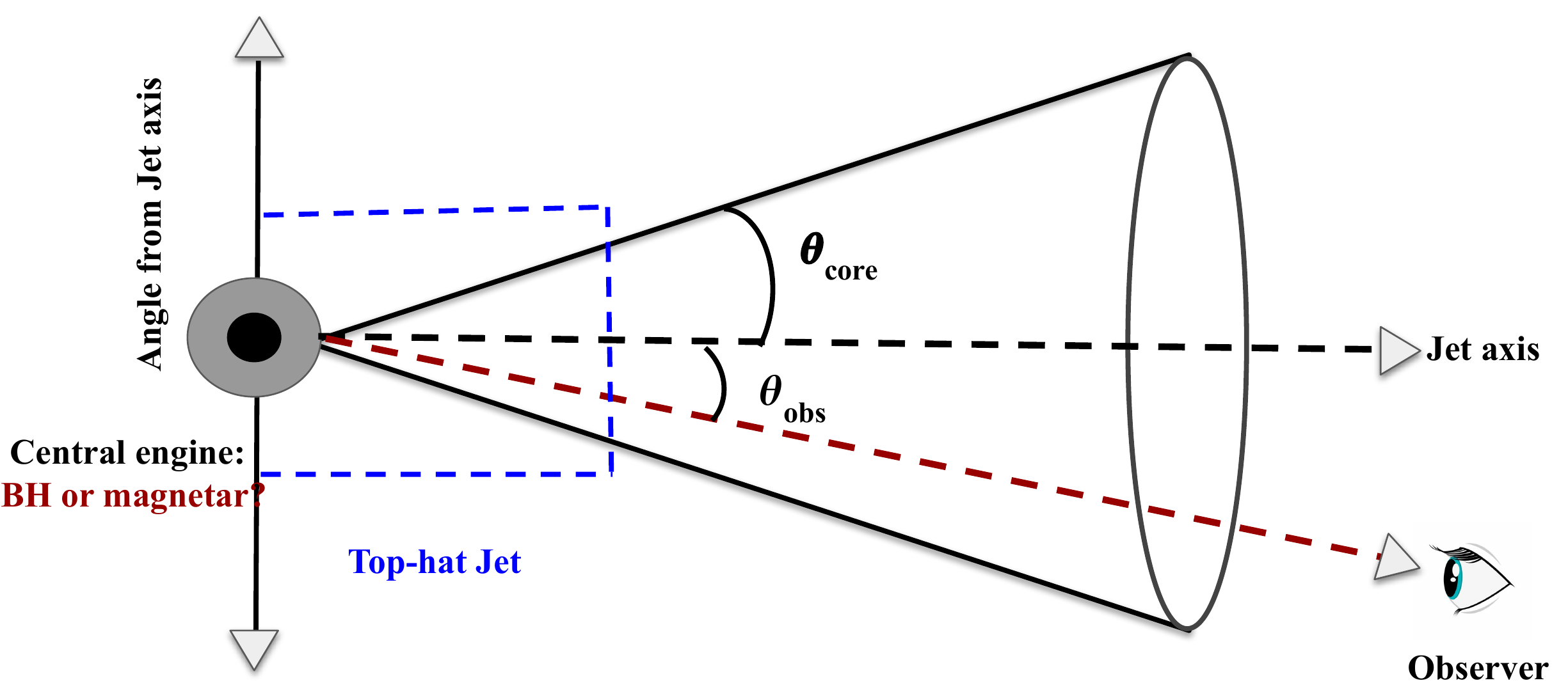}
\caption{A schematic diagram of a top-hat jet of a GRB for the on-axis view ($\theta_{obs}$ $<$ $\theta_{core}$) is presented, one of the possible scenario for ZTF21aaeyldq.}
\label{jet}
\end{figure}

We noticed that the temporal evolution of optical data of ZTF21aaeyldq is consistent with that of typical GRB afterglows. The jet break signature in the optical data indicates that the source observing angle was within the jet opening angle. There could be three possible reasons for the origin of ZTF21aaeyldq. The simplest explanation for any orphan afterglow is that the source is observed on-axis ($\theta_{obs}$ $<$ $\theta_{core}$), however, the gamma-ray counterpart was unambiguously missed by space-based gamma-ray GRBs missions, either due to their sensitivity or source was not in their field of view. Our afterglow modelling suggests that this could be the case for the ZTF21aaeyldq. A schematic diagram showing a top-hat jet for the on-axis view is presented in Fig. \ref{jet}. Other natural explanations for orphan afterglows could be off-axis observations ($\theta_{obs}$ $>$ $\theta_{core}$) and a dirty fireball.

Some authors use the off-axis jet scenarios to explain the orphan afterglows \citep{2002ApJ...570L..61G}. The off-axis models are also useful to explain the short GRBs from structured jets \citep{2018MNRAS.473L.121K}, low-luminosity bursts \citep{2007ApJ...662.1111L} and the X-ray plateaus present in the X-ray afterglow light curves \citep{2020MNRAS.492.2847B}. However, in the case of ZTF21aaeyldq, our afterglow modelling suggests that ZTF21aaeyldq was a classical burst viewed on-axis, so we discard the off-axis possibility. In addition, \cite{2020ApJ...896..166R} suggests a few other signatures of off-axis observations, such as an early shallow decay and a large value of change in temporal indices before and after the break ($\Delta \alpha$). However, in the case of ZTF21aaeyldq, we did not notice an early shallow decay, and also $\Delta \alpha$ value is not very large. Therefore, we discard the possibility of off-axis observations.

The third possibility to explain the orphan afterglow is a dirty fireball, i.e., fireball ejecta with a low value of bulk Lorentz factor ($\Gamma_0 \sim$ 5) and with large numbers of baryons. For a dirty fireball, gamma-ray photons of prompt emission are absorbed due to the optical thickness. In such a case, we expect a large value of deceleration time due to a low value of $\Gamma_0$. The deceleration time for a constant density medium is defined as: $T_{\rm dec}$ = 59.6 $\times ~ n_{0}^{-1/3}$ $\Gamma_{0, 2.5}^{-8/3}$  $E_{k, 55}^{1/3}$ s \citep{2021MNRAS.505.1718J}. Considering $\Gamma_0$ equal to 100, we calculated $T_{\rm dec}$ = 210.68 s using the parameters obtained from afterglow modelling, but for $\Gamma_0$ equal to 10, we calculated $T_{\rm dec}$ = 1.13 days. We looked for the GCNs circular to find the bursts within two days before the ZTF discovery of ZTF21aaeyldq, but we did not found any coincident GRBs. In addition, we constrain the limit on bulk Lorentz factor ($\Gamma_{0}$ $<$ 202) using the prompt emission correlation between $\Gamma_{0}$-$E_{\gamma, \rm iso}$\footnote{$\Gamma_{0}$ $\approx$ 182 $\times$ $E_{\gamma, \rm iso, 52}^{0.25}$} \citep{2010ApJ...725.2209L}. In the light of the above discussion, we discard the possibility of a dirty fireball.

\section{Summary and Conclusion}
\label{Summary and Conclusion}

Optical follow-up observations are helpful to constrain the jet geometry, environment, total energy, and other microphysical parameters of GRBs afterglows. India has a longitudinal advantage and a long history for these ToO observations. The recent installation of 3.6m DOT at Devasthal observatory of ARIES help us to move one step forward to these scientific goals. The back-end instruments of DOT are crucial for deep imaging and spectroscopy of afterglows. In this work, we present the analysis of two exciting afterglow sources. For GRB 210205A, an X-ray afterglow was detected by \swift XRT, but we do not find any optical counterpart despite deep follow-up observations using the 3.6m DOT telescope.
We investigate the possible reason for the optical darkness of the source. We noticed that GRB 210205A has a fainter X-ray emission compared to well-known X-ray afterglows detected by \swift XRT telescope. We obtained the optical-to-X-ray spectral index and compared it with a large sample of GRBs. We found a hint that the source satisfied the recent definition of the darkness of the afterglow. We also measure the host galaxy extinction using afterglow SED, suggesting either intrinsic faintness of the source or high redshift origin as a possible reason for the optical darkness of the afterglow.


In the case of ZTF21aaeyldq, no gamma-ray prompt emission counterpart was reported by any space-based satellites, and the ZTF survey discovered this source. Based on optical identification of the afterglow, XRT observations and other follow-up observations found a fading source characterized with typical afterglow nature with redshift measured using the 10.4m GTC. Following key observational facts confirms the afterglow nature of ZTF21aaeyldq. (i) Fading nature of the source consistent with typical temporal index seen in case of afterglows. (ii) jet break in the optical light curve (iii) The detection of X-ray afterglow with typical luminosity. We performed detailed multiwavelength modelling of the afterglow, including data taken with 3.6m DOT. Our results suggest that afterglow could be described with a slow cooling case for an ISM-like ambient medium. The micro-physical afterglow parameters of ZTF21aaeyldq are consistent with the typical afterglow parameters of other well-studied GRBs. Our modelling also suggests that ZTF21aaeyldq was viewed on-axis ($\theta_{obs}$ $<$ $\theta_{core}$), however, the gamma-ray counterpart was unambiguously missed by space-based gamma-ray GRBs missions, either due to their sensitivity (weak prompt emission) or source was not in their field of view (Earth occultation). It helps us rule out other hypotheses for the possible origin of orphan afterglows, such as off-axis view and a low-Lorentz factor jet. We also compared the light curve evolution of ZTF21aaeyldq with other known orphan afterglows with a measured redshift. We found that the nature of the afterglow of ZTF21aaeyldq is very similar to other well-known orphan afterglows ZTF20aajnksq and ZTF19abvizsw. The comparison also indicates that during the late epochs, ZTF21aaeyldq was fainter than other known cases of orphan afterglows. Further, we measure the energetics of prompt emission (using the limit on $E_{\rm \gamma, iso}$ $\leq 1.52 \times 10^{52}$ erg) and afterglow phases of ZTF21aaeyldq and estimated a radiative efficiency ($\eta$= $E_{\rm \gamma, iso}$/($E_{\rm \gamma, iso}$+ $E_{k}$)) $\leq$ 28.6 \%. This suggests that most energy is used as kinetic energy and supports the internal shocks as the likely origin of prompt emission. We expect to detect more similar sources in the current era of survey telescopes with large fields of view like ZTF and others. Our results also highlight that the 3.6m DOT has a unique capability for deep follow-up observations of similar and new transients to deeper limits as a part of time-domain astronomy.

\section*{Acknowledgements}
We thank the anonymous referee for providing positive and constructive comments to improve the manuscript. RG, SBP, and KM acknowledge BRICS grant {DST/IMRCD/BRICS/PilotCall1/ProFCheap/2017(G)} for the financial support. AA acknowledges funds and assistance provided by the Council of Scientific \& Industrial Research (CSIR), India with file no. 09/948(0003)/2020-EMR-I. This research is based on observations obtained at the 3.6m Devasthal Optical Telescope (DOT) during observing cycles DOT-2021-C1 and DOT-2020-C2, which is a National Facility run and managed by Aryabhatta Research Institute of Observational Sciences (ARIES), an autonomous Institute under Department of Science and Technology, Government of India. Based on observations collected at Centro Astronómico Hispano-Alemán (CAHA) at Calar Alto, operated jointly by Junta de Andalucía and Consejo Superior de Investigaciones Científicas (IAA-CSIC). This research has made use of data obtained from the High Energy Astrophysics Science Archive Research Center (HEASARC) and the Leicester Database and Archive Service (LEDAS), provided by NASA's Goddard Space Flight Center and the Department of Physics and Astronomy, Leicester University, UK, respectively.

\vspace{-1em}


\end{document}